\documentclass[reprint,amsmath,amssymb,aps]{revtex4-2}

\usepackage{graphicx}
\usepackage{dcolumn}
\usepackage{bm}
\usepackage[dvipsnames]{xcolor}
\usepackage{cancel}

\begin{document}

\preprint{APS/123-QED}
\title{Nonreciprocity in cavity magnonics at millikelvin temperature}
\author{Mun Kim}
\author{Armin Tabesh}
\author{Tyler Zegray}
\author{Shabir Barzanjeh}
    \homepage{https://quantumcir.com/}
\author{Can-Ming Hu}
    \homepage{http://www2.physics.umanitoba.ca/u/hu/}
    \affiliation{Department of Physics and Astronomy, University of Manitoba, Winnipeg, Canada R3T 2N2}
    \affiliation{Institute for Quantum Science and Technology (IQST),
University of Calgary, Calgary, Canada T2N 1N4}

\date{\today}

\begin{abstract}
Incorporating cavity magnonics has opened up a new avenue in controlling non-reciprocity. This work examines a yttrium iron garnet sphere coupled to a planar microwave cavity at milli-Kelvin temperature. Non-reciprocal device behavior results from the cooperation of coherent and dissipative coupling between the Kittel mode and a microwave cavity mode. The device's bi-directional transmission was measured and compared to the theory derived previously in the room temperature experiment. Investigations are also conducted into key performance metrics such as isolation, bandwidth, and insertion loss. The findings point to the coexistence of coherent and dissipative interactions at cryogenic conditions, and one can leverage their cooperation to achieve directional isolation. This work foreshadows the application of a cavity magnonic isolator for on-chip readout and signal processing in superconducting circuitry.

\end{abstract}

\maketitle

\section{Introduction}
Numerous technologies working at microwave frequency, including test and measurement circuits, simultaneous transmit-and-receive architecture, and wireless transmission, require non-reciprocal components such as isolators and circulators. Their function in the classical regime is either to protect delicate components from high power reflections or to route the outgoing and incoming signals to the appropriate transmitter and receiver \cite{989957}. Meanwhile, isolators and circulators are also employed in cryogenic quantum mechanical experiments to shield sensitive signals from backscattering noise \cite{8661687,walter2017rapid,vijay2011observation,hatridge2013quantum,vijay2011observation}.

Traditionally, non-reciprocal components are designed using ferrite materials, which lose their Lorentz reciprocity under the application of an external magnetic field \cite{lax1963microwave}. One of the widely used configurations is the stripline Y-junction type \cite{1125923}. The design is renowned for its power handling ability (tens to hundreds of Watts) and incredibly low loss ($<1$ dB). The isolation and bandwidth are, however, limited to a point \cite{1125244}. Consequently, devices are cascaded in some applications to reach the desired performance. Because of this, there has been a significant effort over the past decades on alternative non-reciprocal technology to engineer cost-efficient and small-form-factor devices with improved performance. 

On the one hand, there are magnet free approaches that can be integrated into a chip-scale structure while providing considerable port isolation ($>20$ dB). Transistors and temporal modulation are the two categories \cite{42266,58677,8661642,nagulu2020non,bi2011chip}. Even though the former is readily compatible with MMIC (Monolithic Microwave Integrated Circuit) and can be realized within a compact package, the transistor's added noise and nonlinear distortion have prevented them from being widely deployed \cite{821785}. Reciprocity can also be violated through spatiotemporally modulated waveguides \cite{lira2012electrically,zanjani2014one,nagulu2020non}. In practice, the material being modulated is typically associated with considerable insertion loss at higher frequency regime. Parallel to these, the rapidly expanding optomechanical and electromechanical systems have shown great potential with lately discovered isolation and circulation effect \cite{shen2016experimental,ruesink2016nonreciprocity,fang2017generalized,barzanjeh2017mechanical}.

On the other hand, coupling magnons (collective excitation of the spin ensemble) with the microwave cavity can produce non-reciprocity. Both circulators and isolators have been realized recently \cite{zhu2020magnon, zhang2020broadband, PhysRevLett.123.127202, shi2021mirror}. In a coherently coupled system, the magnon mode can strongly couple with a selected chiral cavity mode to produce an asymmetrical transmission profile that spans nearly $500$ MHz \cite{zhang2020broadband}. A circulator utilizing the similar scheme achieved over 50 dB of port isolation \cite{zhu2020magnon}. Alternatively, the cooperation of coherent and dissipative coupling also produces non-reciprocity that offers complete isolation ($>80$ dB) \cite{PhysRevLett.123.127202}. The eigenmodes in such a system couple differently with the microwave traveling in the opposite direction. In addition, the repulsive behavior of linewidth is exploited to fully compensate for the hybridized mode's damping. A sharp unidirectional rejection band develops as a result. The working principle is later found applicable to a circulator \cite{shi2021mirror}. 

Considering the current ferrite non-reciprocal devices occupy substantial space inside the dilution refrigerator and hence limiting the number of qubits to be incorporated, the demand for compact and effective circulators and isolators still remains \cite{ranzani2019circulators}. Cavity magnonic's tunability \cite{rameshti2022cavity, harder2021coherent}, combined with recent efforts to push toward on-chip integration, has proven to be beneficial for device design \cite{li2020hybrid}. The nearly perfect isolation found in \cite{PhysRevLett.123.127202} is also attractive for sensitive signal detection.  While the coherent coupling has been proven to exist throughout a wide temperature range, \cite{PhysRevB.97.184420,PhysRevLett.111.127003, PhysRevLett.113.083603}, it is yet to be validated if the same is true for dissipative coupling. Hereof, this work examines the bi-directional transmission coefficient of a cavity magnonic system at a millikelvin temperature in search of non-reciprocity.

\section{Theoretical Framework}
\begin{figure}[t]
\includegraphics[trim = 13mm 5mm 15mm 10mm, clip, width=8.5cm]{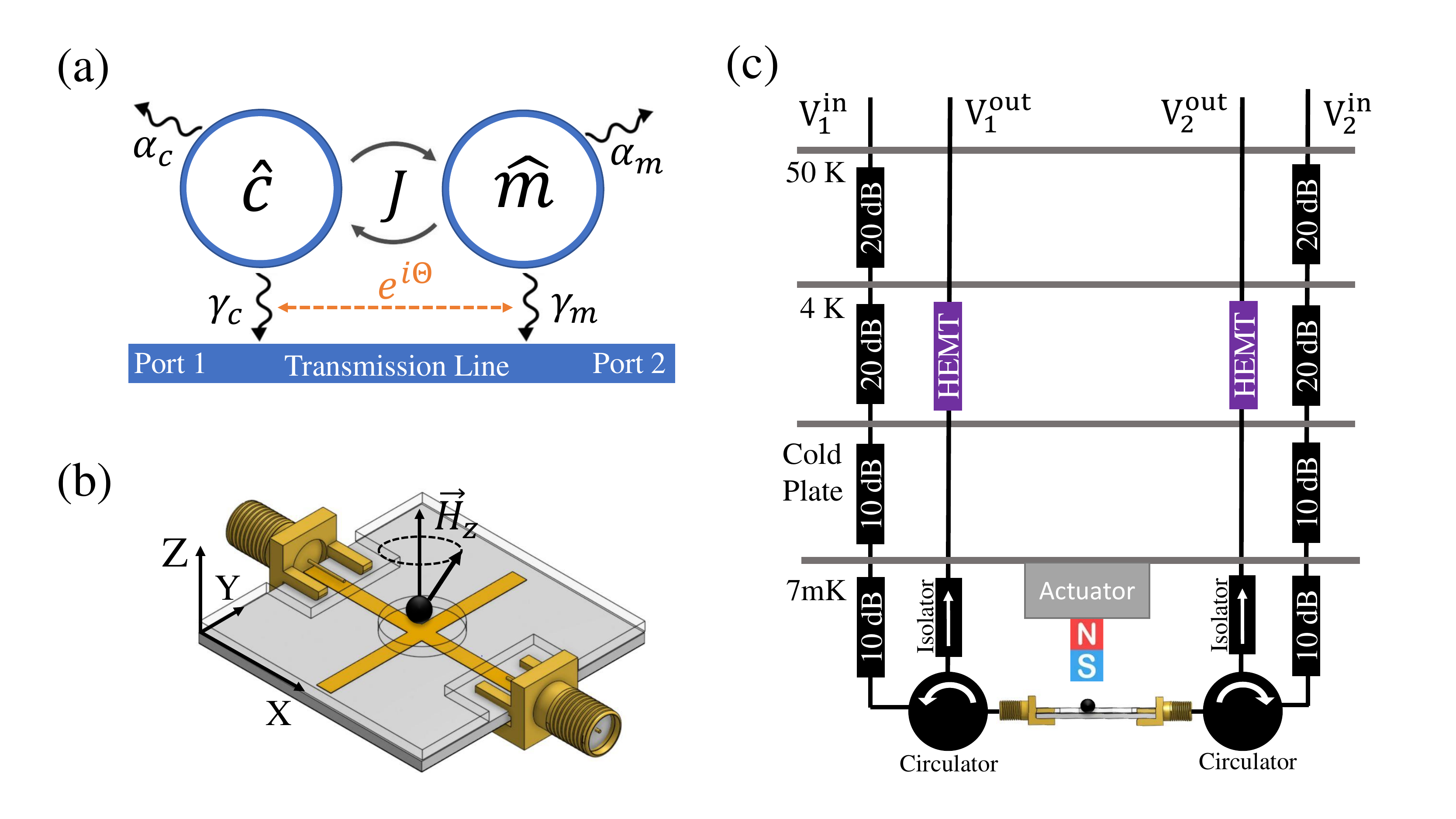}\\
\caption{\label{fig1} (a) Schematic diagram showing interactions inside a cavity magnonic system. (b) The planar microwave circuit supports both standing and travelling waves. The YIG sphere is cemented onto the sample holder and placed nearby the cross junction. (c) Schematic of the measurement apparatus. The cavity magnonic assembly is cooled to 7mK inside a dilution fridge. A tunable static magnetic field  is applied perpendicularly to the cavity board by the actuator. Forward and backward transmission spectra are obtained through scalar network analysis. i.e. $|S_{21(12)}|^2=20\,\log_{10}(V_{1(2)}^{\textrm{in}}/V_{2(1)}^{\textrm{out}})$  }
\end{figure}

There are two different types of interactions for an open cavity magnonic system [Fig. \ref{fig1}(a)]. The spatial overlap of the magnon mode with the cavity microwave field leads to the coherent coupling \cite{PhysRevLett.113.156401}. The mutual dissipative coupling exists as both the cavity and magnon modes simultaneously discharge energy to the transmission line. In this regard, the Hamiltonian after rotating wave approximation (RWA) takes the form of \cite{PhysRevLett.123.127202}
\begin{equation}
\hat{H}/\hbar=\tilde{\omega}_c\hat{c}^\dagger \hat{c}+\tilde{\omega}_m\hat{m}^\dagger\hat{m}+(J-i\Gamma e^{i\Theta_{1(2)}})(\hat{c}^\dagger\hat{m}+\hat{m}^\dagger\hat{c})
\label{eq.1}
\end{equation} 
\noindent where $\hat{c}^\dagger(\hat{c})$ and $\hat{m}^\dagger(\hat{m})$ are the creation (annihilation) operators of the cavity and magnon modes, respectively. Their uncoupled complex frequencies are specified as $\tilde{\omega}_{c,m}=\omega_{c,m}-i\alpha_{c,m}$, with $\alpha_{c,m}$ being the corresponding intrinsic damping rates. Regarding the last term, the lossless energy exchange process between the cavity and magnon modes is quantified by the real coherent coupling strength $J$. The indirect, reservoir-mediated interaction between two subsystems is described by $i\Gamma$. Often known as the dissipative coupling strength, this imaginary term is related to the extrinsic damping rates of cavity ($\gamma_c$) and magnon modes ($\gamma_m$), i.e., $\Gamma=\sqrt{\gamma_c\gamma_m}$. In addition, $\Theta_{1(2)}=0 \, \textrm{or}\, \pi$ such that $e^{i\Theta_{1(2)}}$ establishes a sign convention for the $\Gamma$ term with respect to the direction of traveling wave. 

The hybridized frequencies are governed by
\begin{equation}
\begin{split}
\tilde{\omega}_{\pm} &= \frac{1}{2} \Bigg[ \omega_c+\omega_m-i(\alpha_c+\alpha_m) \\
&\pm\sqrt{[(\omega_c-\omega_m)-i(\alpha_c-\alpha_m)]^2+4(J-i\Gamma e^{i\Theta_{1(2)}})^2}   \Bigg]  
\end{split}
\label{eq.2}
\end{equation}
\indent The real part of $\tilde{\omega}_\pm$ portrays the dispersion, whereas the imaginary part controls linewidth behavior. By carefully examining the Eq. (\ref{eq.2}), it can be seen that $\operatorname{Re}(\tilde{\omega}_{\pm})$ are the same regardless of the chosen $\Theta_{1(2)}$ value. Thus, the system exhibits an identical dispersion for both forward and backward transmission measurements. In contrast, the imaginary part at the $(J-i\Gamma e^{i\Theta})^2$ term is different, implying that the energy dissipation of hybridized modes depends on the measurement direction.

\indent The scattering parameters of the system are found to be \cite{PhysRevLett.123.127202} 
\begin{equation}
S_{21,(12)} = 1+\frac{\gamma_c}{[i(\omega-\omega_c)-(\alpha_c+\gamma_c)]-\frac{[iJ+\Gamma e^{i\Theta_{1(2)}}]^2}{i(\omega-\omega_m)-(\alpha_m+\gamma_m)}} 
\label{eq.s21}
\end{equation}
where $\Theta_1$ and $\Theta_2$ are associated with $S_{21}$ and $S_{12}$, respectively.  Notably, the $\pm2iJ\Gamma$ term produces asymmetrical transmission profile, $|S_{21}|^2\neq |S_{12}|^2$. This direction dependent cooperative term requires the magnetic field of the traveling wave and cavity to interfere, so that spin precession is favored in one measurement direction while suppressed in the other, i.e., the chirality of spin precession allows the magnon mode to couple differently to the microwave field with opposite helicity. Whereas in the case of a pure interaction ($J=0$ or $\Gamma=0$), the s-parameters are reciprocal.

\section{Experiment}

\begin{figure}[t]
\includegraphics[trim = 6mm 0mm 2mm 12mm, clip, width=8.5cm]{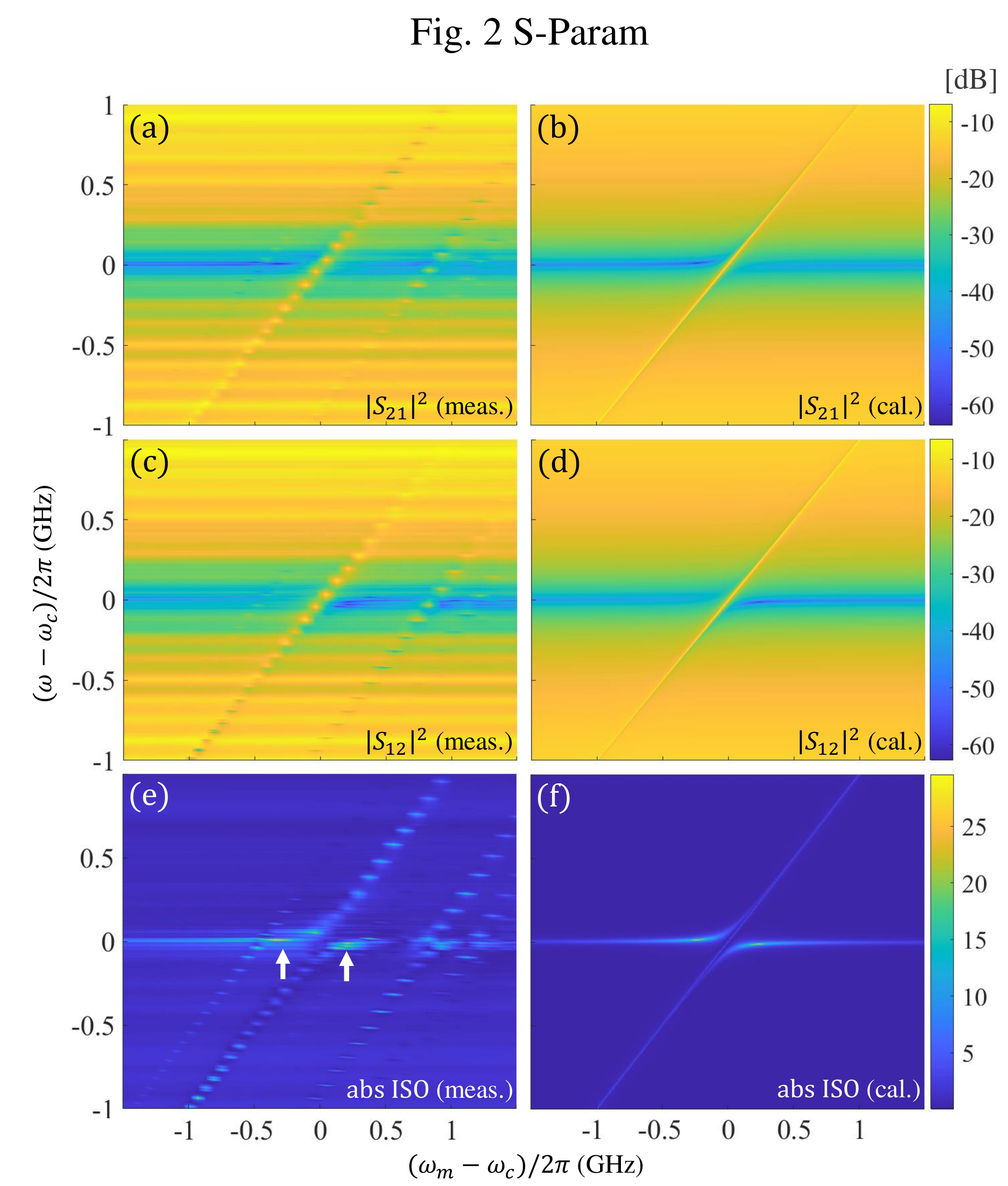}\\
\caption{\label{fig2} Transmission profile of the nonreciprocal device. (a) Forward transmission amplitude ($|S_{21}|^2$) as a function of frequency detuning and field detuning. The probing microwave power is 0 dBm. (c) Backward transmission ($|S_{12}|^2$) shows similar behavior as in (a). (e) The absolute difference between forward and backward transmission. Isolation peaks twice near zero field detuning (b),(d) and (f) are the corresponding calculation results.}
\end{figure}

\subsection{Experimental Setup}
A single yttrium iron garnet (YIG) sphere placed close to the intersection of a planar cross-shaped microwave cavity makes up the device under test (DUT). The sample holder, depicted in Fig. \ref{fig1} (b) as a transparent box, fixes the location of the YIG. A permanent magnet is fastened to the end of an actuator to control the magnetic field $\vec{H}_z$. In this way, the ferrite is assumed to be saturated, and $\omega_m$ can be tuned by varying the actuator's position. The entire setup is housed inside a dilution refrigerator and the sample is anchored to the mixing plate of the refrigerator, as shown in Fig.\ref{fig1} (c).

The YIG sphere, with a diameter of $1$ mm was provided by Ferrisphere. Inc \cite{ferrite}. The Kittel mode is the focus of this work since it not only has the highest number of spins coupled with the system's photons, but also has a straightforward dispersion. i.e., $\omega_m=\gamma(\vec{H}_A+\vec{H}_z)$, where $\gamma=2\pi\times26.3\,\mu_0\,\textrm{GHz/T}$ is the gyromagnetic ratio, $\vec{H}_z$ is the static bias magnetic field perpendicular to the cavity board and $\vec{H}_A$ is the magnetocrystalline anisotropy field. During the experiment, the magnon mode's frequency is tuned from $4.57\,\textrm{GHz}$ to $8.89\,\textrm{GHz}$, and the intrinsic damping rate is found to be $\alpha_m/2\pi=1\,\textrm{MHz}$.  

A ROGERS RO4003C $(\epsilon_r=3.55)$ laminate measuring 0.81 mm thick is used to construct the microwave cavity. The 50$\,\Omega$ transmission line that connects the two ports allows for bi-directional transmission measurements. Its midpoint joins two open-ended 6.6 mm long stubs functioning as half-wavelength resonators. A direction-dependent chirality is produced nearby the cross junction as a result of the interaction between the magnetic fields of the transmission line and the stub resonator. The cavity resonant frequency, intrinsic damping rate at millikelvin are $\omega_c/2\pi = 6.354\,\textrm{GHz}, \alpha_c/2\pi=11 \,\textrm{MHz}$, respectively (see Appendix. A).

The coherent interaction occurs because the YIG sphere is located inside the magnetic field of the stub resonator. An extrinsic damping rate of $\gamma_c/2\pi=3\, \textrm{GHz}$ results from the electrical connection between the stub and transmission line. Given the magnon mode's proximity to the transmission line, there also exists a dissipation channel of a similar kind at rate $\gamma_m/2\pi=0.16\, \textrm{MHz}$. It is also important to note that both coupling strengths are maintained constant throughout the experiment due to the fixed YIG position \cite{yang2019control}. Magnon mode frequency is the only tuned variable.

\subsection{Bi-directional Transmission Measurements}

\begin{figure}[t]
\includegraphics[trim = 1mm 24mm 3mm 17mm, clip, width=8.5cm]{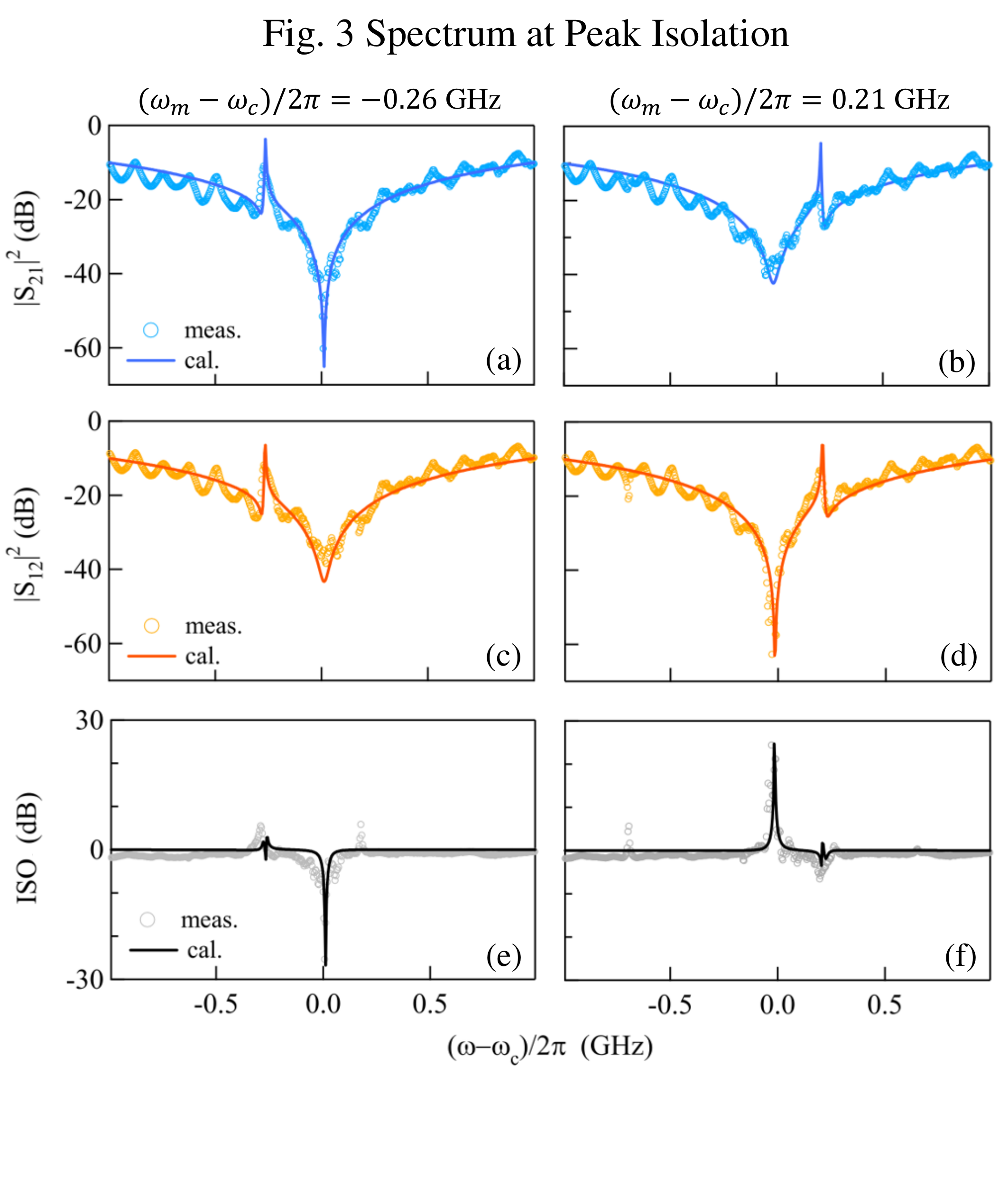}
\caption{\label{fig3} Transmission spectra corresponding to maximum isolation. (a) $|S_{21}|^2$, (c) $|S_{12}|^2$ and (e) $\textrm{ISO}$ obtained at the field detuning of $-0.26\,\textrm{GHz}$. Attenuation on the forward traveling wave is 26.9 dB.  (b) $|S_{21}|^2$, (d) $|S_{12}|^2$ and (f) $\textrm{ISO}$ obtained at the field detuning of $0.21\,\textrm{GHz}$. Attenuation on the backward traveling wave is 25.9 dB. }
\end{figure}

Two-way transmission profiles of the DUT are measured first, then their difference is calculated. The measured forward and backward transmission coefficients $|S_{21,12}|^2$ are shown in Figs. \ref{fig2}(a) and (c) as a function of frequency detuning: $ (\omega-\omega_c)/2\pi$, and field detuning: $(\omega_m-\omega_c)/2\pi$. At the degeneracy point, a normal mode splitting of approximately 110 MHz is present in both directions, orders of magnitude wider than the intrinsic damping rates, indicating the magnon mode is strongly coupled to the cavity mode. 

Several obscure anti-crossings are found near $(\omega_m-\omega_c)/2\pi=-0.5\,\textrm{GHz and}\,1.0\,\textrm{GHz}$, which can be related to the weak coupling between the cavity mode and other magnetostatic modes induced by inhomogeneity of magnetic fields. They are not considered in the scope of this work.

Next, the coupling strengths are evaluated by fitting Eq. (\ref{eq.s21}) to each measured transmission spectra. The average of coherent coupling strength is $J/2\pi=63\,\textrm{MHz}$, whereas $\Gamma/2\pi=22\,\textrm{MHz}$ for the dissipative. $\Theta$ takes $\pi$ for $|S_{21}|^2$ and $0$ for $|S_{12}|^2$. A good agreement between the fitting results and the raw measurement data is seen in Fig. \ref{fig2}(b) and (d). 

One notices that the mode splitting, being 110 $\,\textrm{MHz}$, is less than that of a pure coherent case $2J/2\pi=126\,\textrm{MHz}$ \cite{harder2021coherent}. It is in fact, altered by the dissipative interaction, which causes energy level attraction instead \cite{harder2018level}. Nevertheless, the transmission mapping still exhibits anti-crossing behavior since $J>\Gamma$. 

Fig. \ref{fig2} (e) and (f) display the absolute value of isolation ($\textrm{ISO}=|S_{21}|^2-|S_{12}|^2$). Two bright peaks at the field detuning $(\omega_m-\omega_c)/2\pi=-0.26, 0.21\,\textrm{GHz}$ [white arrows] suggest a non-reciprocal transmission, further confirming that two subsystems are not interacting in a purely coherent manner.

\subsection{Non-reciprocity Characterization}

\begin{figure}[t]
\includegraphics[trim = 2mm 22mm 12mm 17mm, clip, width=8.5cm]{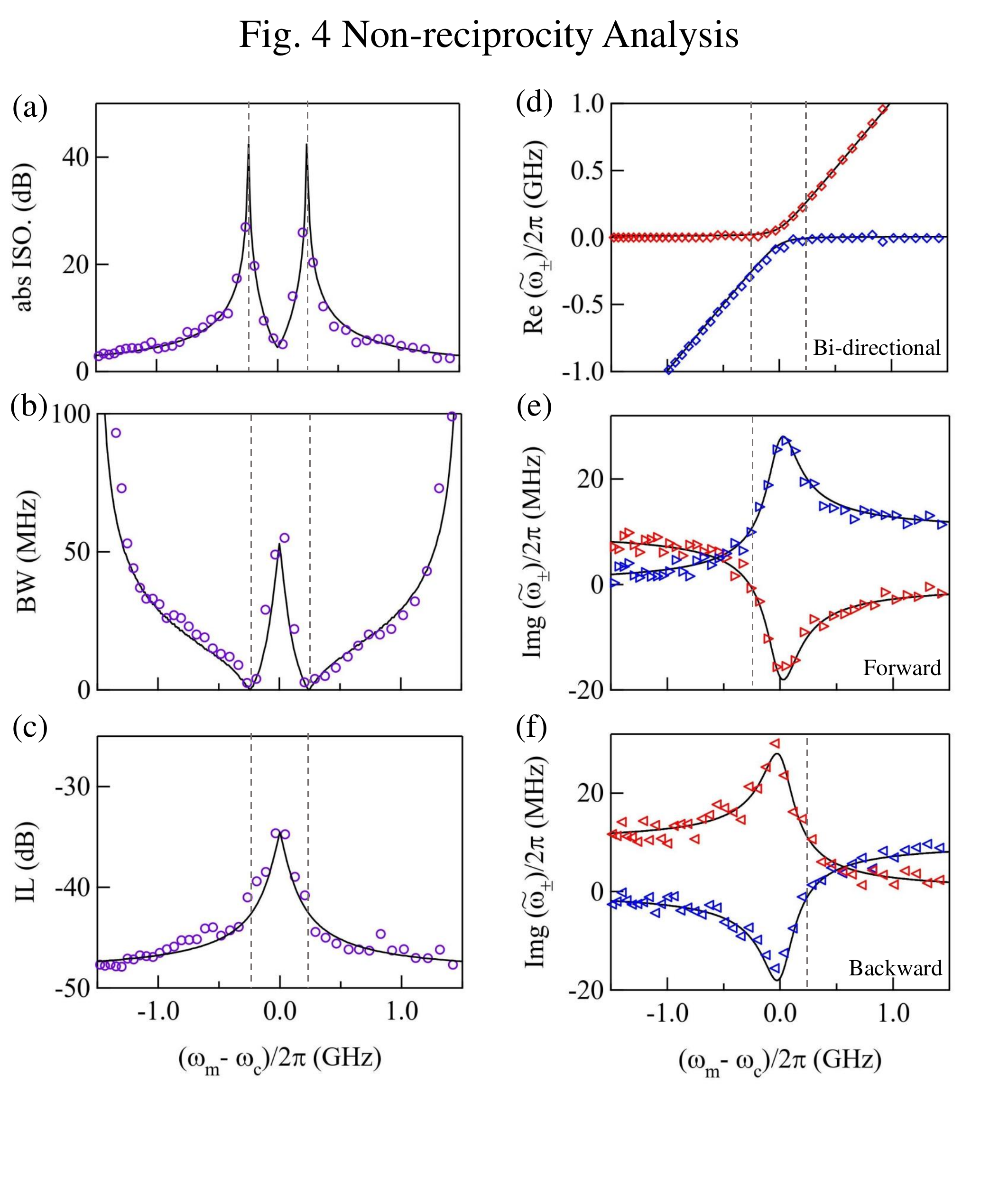}
\caption{\label{fig4} (a),(b) and (c) Absolute value of isolation, bandwidth and insertion loss against frequency detunning. Two zero damping conditions are marked by dashed lines. There exists a trade off between isolation and bandwidth, but not insertion loss. (d) Real part of $\tilde{\omega}_\pm$ against field detuning. (e) and (f) Imaginary part of $\tilde{\omega}_\pm$ corresponding to forward and backward traveling waves. Dashed lines mark where the hybridized mode's linewidth crosses zero.}
\end{figure}

The forward and backward transmission shown in Fig. \ref{fig3}(a),(c) and (b),(d) correspond to the two isolation peaks identified previously. According to Fig.\ref{fig3} (e) and (f), the respective isolation is $-26.9\,\textrm{dB}$ and $25.9\,\textrm{dB}$,  emerging at frequency detuning conditions $0.012\, \textrm {GHz}$ and $-0.016\, \textrm {GHz}$.

Due to inadequate calibration on the feed and return signal path [$V^{\textrm{in/out}}_{1,2}$ in Fig. \ref{fig1}(c)], there is an imperceptible offset between forward and backward measurements, with an average of $1.49\,\textrm{dB}$ across the sweep frequency span [Fig. 5 in Appendix.C]. Nevertheless, the isolation is substantially greater than the offset; The DUT is indeed non-reciprocal. 

For practical applications, the performance characteristics of a non-reciprocal device, such as isolation, insertion loss (IL), and bandwidth (BW), are examined across the range of field detuning.

As depicted in Fig. \ref{fig4} (a), the theoretical isolation is maximum at $(\omega_m-\omega_c)/2\pi=\pm0.23\,\textrm{GHz}$, implying mirror symmetric non-reciprocity with regard to the cavity mode frequency. The attenuation of the forward traveling wave causes the left peak, whereas the attenuation of the backward traveling wave results in the right peak. Away from the detuning conditions mentioned above, the isolation drastically decreases. The device is anticipated to exhibit even larger isolation if the tuning resolution of the magnetic field can be improved.

The 3 dB bandwidth is shown in Fig. \ref{fig4}(b). The width is obtained by counting up $3$ dB from the transmission minimum. When comparing Fig. \ref{fig4}(b) to (a), a trade-off between the isolation and bandwidth becomes obvious. Maximum isolation is achieved with minimum bandwidth at $(\omega_m-\omega_c)/2\pi=\pm0.23\,\textrm{GHz}$ (marked by dashed lines). This is consistent with previously discovered zero damping condition (ZDC) at room temperature \cite{PhysRevLett.113.156401}. From the oscillator's point of view, a zero bandwidth oscillator absorbs all the energy from the transmission line. The two isolation peaks found in measurement correspond to bandwidths of 2.5 and 2.7 MHz, respectively. Once the field detuning surpasses $\pm1.3$ GHz, the isolation becomes less than 3 dB, and the bandwidth is indeterminate.

Fig. \ref{fig4}(c) illustrates device insertion loss, defined as $\textrm{IL}=|S_{12}|^2$ for negative field detuning and $\textrm{IL}=|S_{21}|^2$ for positive field detuning.  Interestingly, there is no apparent trade-off between ISO and IL.  Therefore, the system can be biased nearby, but not at the ZDC for an optimized performance. When the field detuning is large, system is in the dispersive regime. Consequently, the insertion loss is an approximate to the cavity's resonant dip. The reason for reduced insertion loss at lower detuning range is explained below. 

The real part of hybridised frequencies is highlighted in Fig. \ref{fig4}(d), which is extracted by reading off transmission zeros. There is a high degree of agreement with Fig. \ref{fig2}. 
Imaginary parts are presented in Fig. \ref{fig4}(e) and (f). The linewidth is extracted by fitting the reciprocal of power amplitude to a superposition of two Lorentzian curves [See Appendix B]. Notably, the imaginary part of the system depends on the direction of the traveling wave. This is due to the complex coupling strength $J\pm i\Gamma$, whose directional dependence determines the sign of the $i\Gamma$ term. Near the zero field detuning, there is a negative damping regime where attenuation on the travelling wave is weakened. Consequently, two effects will be observed in the resonance dip: Reduced insertion loss and linewidth broadening.

As the final step of the experiment, the travelling wave's power is varied from $10\,\textrm{dBm}$ to $-40\,\textrm{dBm}$. The system parameters are found to be irrelevant of probing microwave power. It is expected here since the number of photons is much smaller than the number of spins such that the system sits inside the linear response regime, and much larger than quantum level phenomenon.

\section{Discussion}
In summary, the performance of an isolator based on cavity magnonic interaction was investigated at millikelvin temperature. The interference between coherent and dissipative coupling causes non-reciprocity. Experimental results confirm earlier discovery \cite{PhysRevLett.123.127202} applies to all temperature ranges. The cryogenic performance of such a cavity magnonic isolator can be tuned and improved in the same way as the room temperature experiments have demonstrated \cite{kim2022prototyping}.

This work has two additional implications. One is that at low temperatures, dissipative coupling exists. Creating novel hybrid systems for quantum information science may benefit from such validation. Second, normal mode splitting is impacted by dissipative coupling. The ferrite material may interact with the travelling wave when cavity magnonic platforms develop a more compact profile. Thus, evaluating the forward and backward transmission is necessary for better calibrated coupling strength.

\begin{acknowledgments}
This work has been funded by NSERC Discovery Grants and NSERC Discovery Accelerator Supplements (C.-M. H.). We would like to thank Dr. Yongsheng Gui for discussions. S.B. acknowledges funding by the Natural Sciences and Engineering Research Council of Canada (NSERC) through its Discovery Grant, funding and advisory support provided by Alberta Innovates through the Accelerating Innovations into CarE (AICE) -- Concepts Program, and support from Alberta Innovates and NSERC through Advance Grant. This project is funded [in part] by the Government of Canada.
\end{acknowledgments}

\appendix

\section{Unloaded Cavity}
The transmission coefficient of the microwave cavity in the absence of ferrite loading can be calculated from Eq. (\ref{eq.s21}) by eliminating both the magnon and coupling terms 
\begin{equation}
S_{21,12}= 1+\frac{\gamma_c}{i(\omega-\omega_c)-(\alpha_c+\gamma_c)}
\label{A2}
\end{equation}
where each variables have been defined previously in the main test.

\indent Next, the cavity is characterized by fitting Eq. (\ref{A2}) to the measured spectrum.  The transmission minimum is located at $\omega_c/2\pi=6.354\,\textrm{GHz}$, the $3\,\textrm{dB}$ bandwidth is the intrinsic damping rate ($\alpha_c/2\pi=11\,\textrm{MHz}$), and  $\gamma_c/2\pi=3\,\textrm{GHz}$ governs the width of the broad background.\\
\begin{figure} 
\includegraphics[trim = 30mm 243mm 30mm 0mm, clip, width=6.5cm]{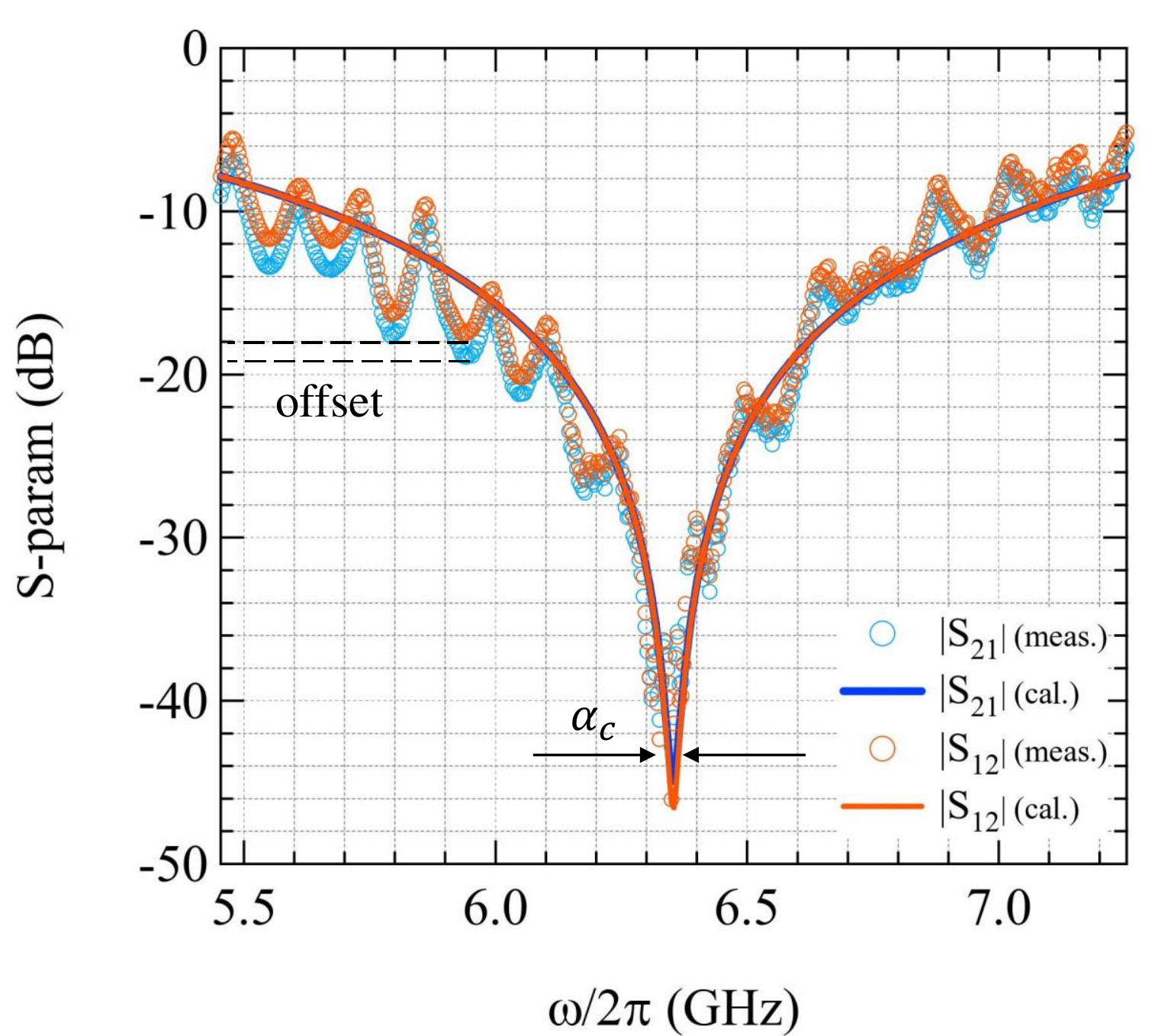}
\caption{ Transmission spectra of the cavity without YIG. The width enclosed by black arrows is the intrinsic damping rate. Offset between forward ($|S_{21}|^2$) and backward ($|S_{12}|^2$) transmission is indicated by dashed lines. The average value of offset across the sweeping frequency span is 1.49 dB. $\omega_c$ and $\alpha_c$ are directly extracted from the measurement, only $\gamma_c$ is obtained from fitting.}
\end{figure}
\begin{figure*} 
\includegraphics[trim = 37mm 40mm 30mm 50mm, width=15cm]{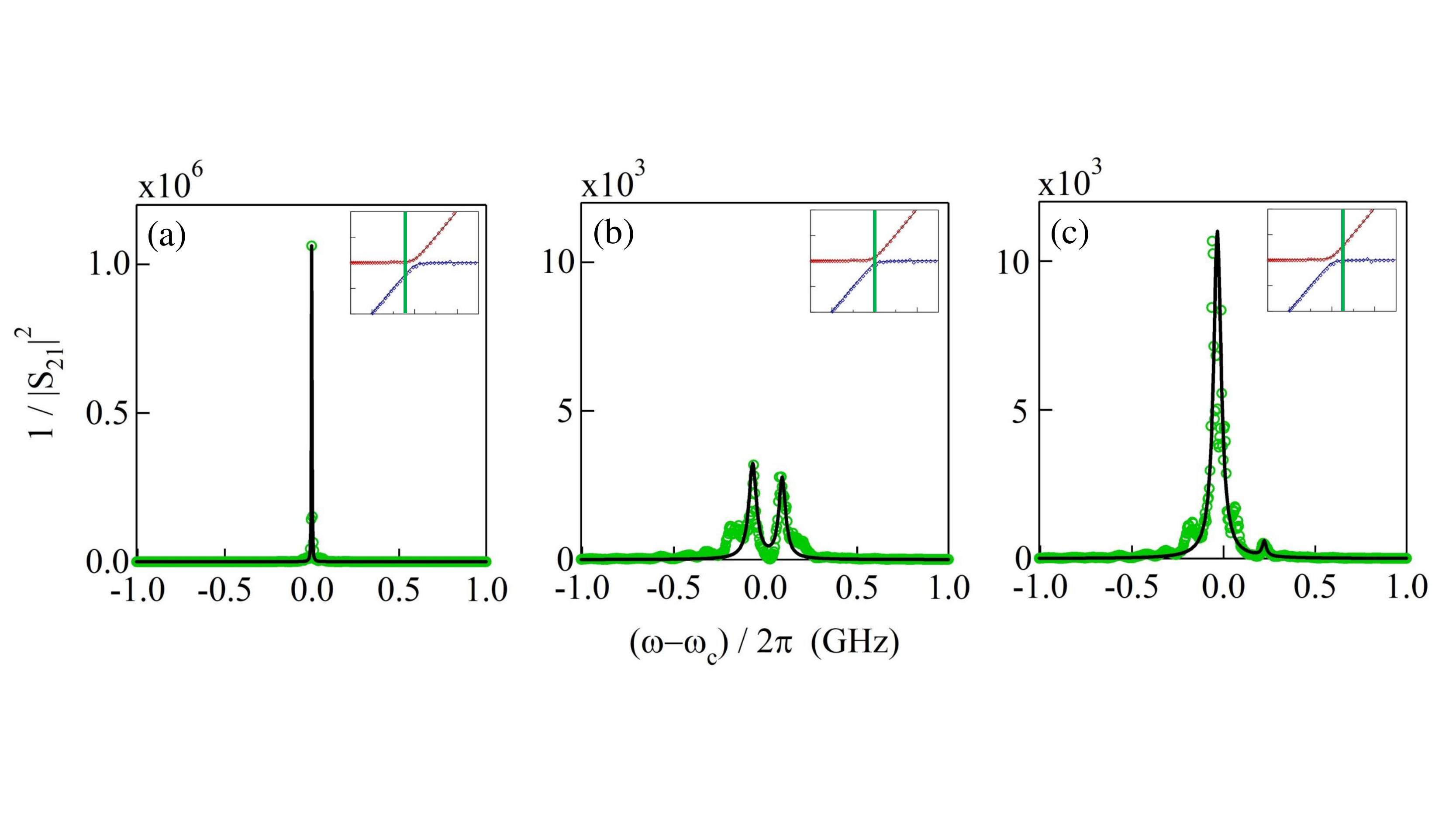}
\caption{ Inverse of power amplitude (green dots) and fitting result (black lines) for forward tranmission spectra. (a),(b) and (c) correspond to field detuning conditions of $(\omega_m-\omega_c)/2\pi = -0.26, 0.04\, \textrm{and}\, 0.21\, \textrm{GHz}$ respectively. List of fitting parameters: (a) $A_{\pm}=1.1\times10^6, 1.0\times10^3$; $\delta\omega_{\pm}=0, 12\,\textrm{MHz}, (\omega-\omega_{\pm})/2\pi =0.01, -0.29\,\textrm{GHz}$. (b) $A_{\pm}=2.7\times10^3, 3.2\times10^3$; $\delta\omega_{\pm}=15, 26\,\textrm{MHz}, (\omega-\omega_{\pm})/2\pi =0.07, -0.10\,\textrm{GHz}$. (c) $A_{\pm}=0.5\times10^3, 10.6\times10^3$; $\delta\omega_{\pm}=10, 20\,\textrm{MHz}, (\omega-\omega_{\pm})/2\pi =0.22, -0.02\,\textrm{GHz}.$ }
\end{figure*}

\section{Fitting Method to Extract $\textrm{Img}(\tilde{\omega}_{\pm})$}
\indent Taking the forward measurement ($\Theta=\pi$), for instance, the transmission coefficient of the system is
\begin{equation}
    S_{21}=1+\frac{\gamma_c}{[i(\omega-\omega_c)-(\alpha_c+\gamma_c)]-\frac{(iJ-\Gamma)^2}{i(\omega-\omega_m)-(\alpha_m+\gamma_m)}}
\end{equation}
\indent Its inverse is
\begin{equation}
\begin{split}
\frac{1}{S_{21}} &=1-\frac{\gamma_c}{[i(\omega-\omega_c)-\alpha_c]-\frac{(iJ-\Gamma)^2}{i(\omega-\omega_m)-(\alpha_m+\gamma_m)}} \\
                & = 1+i\gamma_c\frac{(\omega-\omega_m)+i(\alpha_m+\gamma_m)}{(\omega-\tilde{\omega}_+)(\omega-\tilde{\omega}_-)} 
\end{split}
\label{b2}
\end{equation}

\indent The hybridized frequencies in Eq. \ref{b2}  can be re-written as $\tilde{\omega}_{\pm}=\omega_{\pm}-i\delta\omega_{\pm}$, where $\omega_{\pm}$ and $i\delta\omega_{\pm}$ stand for the Real ($\textrm{Re}(\tilde{\omega}_{\pm})$) and Imaginary ($|\textrm{Img}(\tilde{\omega}_{\pm})|$) parts respectively. The inverse of power amplitude is then found to be proportional to
\begin{equation}
\begin{split}
    \frac{1}{|S_{21}|^2} &=\frac{1}{S_{21}}\cdot\frac{1}{S_{21}^*}\\
    & \propto \gamma_c^2 \frac{(\omega-\omega_m)^2+(\alpha_m+\gamma_m)^2}{[(\omega-\omega_+)^2+\delta\omega_+^2][(\omega-\omega_-)^2+\delta\omega_-^2]}
\end{split}
\end{equation}
which can be decomposed into two Lorentzian lineshapes:
\begin{equation}
\begin{split}
    \frac{1}{|S_{21}|^2} &= \frac{A_+\delta\omega_+^2}{(\omega-\omega_+)^2+\delta\omega_+^2} + \frac{A_-\delta\omega_-^2}{(\omega-\omega_-)^2+\delta\omega_-^2}
\end{split}
\end{equation}
where
\begin{equation}
    A_{\pm}=\bigg ( \frac{\gamma_c^2}{2} \bigg)\bigg( \frac{ (\omega-\omega_m)^2+(\alpha_m+\gamma_m)^2}{[(\omega-\omega_{\mp})^2+\delta\omega_{\mp}^2][\alpha_c+\alpha_m-\delta\omega_{\mp}]^2} \bigg)
\end{equation}
\indent In Eq. (B4), the amplitudes of lineshapes are determined by $A_{\pm}$. As $\omega_{\pm}$ are previously determined by reading off transmission zeros, the only variables remaining to extract are $\delta \omega_{\pm}$. Selected fitting results for forward transmission spectra are presented in Fig. 6.   

According to Eq. (2), the sum of Img($\tilde{\omega}_\pm$) is always a constant. In order words, the total damping of the system should remain unchanged.
\begin{equation}
    \delta(\tilde{\omega}_-) +\delta(\tilde{\omega}_+)=\alpha_c+\alpha_m = 12\, \textrm{MHz}
\end{equation}

However, the $\delta\omega_{\pm}$ obtained from the fitting procedure are all positive, which may exceed 12 MHz. For instance, in Fig. 6(b), $\delta\omega_{\pm}$ correspond to field detuning condition of 0.04 GHz are 26, and 15 MHz, respectively. The resultant total damping is 41 MHz. In this case, $\delta\omega_-$ must take the negative value ( -15 MHz) such that the total damping rates of the system remain consistent.  Similar sign convention can be found in Fig. 6(c), where $\delta\omega_{\pm}=20, -10\,\textrm{MHz}$ is taken for Fig. 4(e).  
\vspace{1cm}

\bibliography{references}% Produces the bibliography via BibTeX.

\end{document}